\documentclass[journal]{IEEEtran}

\usepackage[utf8]{inputenc}
\usepackage{cite,amsmath,amssymb}

\usepackage{graphicx,epstopdf}
\usepackage{epsfig}
\usepackage{textcomp}
\usepackage{xcolor}
\usepackage{hyperref}
\usepackage{cleveref}
\usepackage{orcidlink}
\usepackage{stfloats}
\usepackage{lipsum}
\usepackage{enumitem}
\usepackage{booktabs}

\allowdisplaybreaks
\usepackage[left=1.65cm,right=1.65cm,top=1.72cm,bottom=1.72cm]{geometry}

\IEEEaftertitletext{\vspace{-1.5\baselineskip}}

\begin{document}

\title{Intelligent Spectrum Sharing in Integrated TN-NTNs: A Hierarchical Deep Reinforcement Learning Approach}

\author{%
Muhammad~Umer{~\orcidlink{0009-0001-8751-6100}},
Muhammad~Ahmed~Mohsin{~\orcidlink{0009-0005-2766-0345}},
Ali~Arshad~Nasir{~\orcidlink{0000-0001-5012-1562}},
Hatem~Abou-Zeid{~\orcidlink{0000-0003-4720-5794}}
~and~Syed~Ali~Hassan{~\orcidlink{0000-0002-8572-7377}
}%

\thanks{Muhammad Umer, Muhammad Ahmed Mohsin, and Syed Ali Hassan are with the School of Electrical Engineering and Computer Science (SEECS), National University of Sciences and Technology (NUST), Pakistan.}
\thanks{Ali Arshad Nasir is with the King Fahd University of Petroleum and Minerals (KFUPM), Saudia Arabia.}
\thanks{Hatem Abou-Zeid is with University of Calgary, Canada.}
}

\maketitle

\begin{abstract}
    Integrating non-terrestrial networks (NTNs) with terrestrial networks (TNs) is key to enhancing coverage, capacity, and reliability in future wireless communications. However, the multi-tier, heterogeneous architecture of these integrated TN-NTNs introduces complex challenges in spectrum sharing and interference management. Conventional optimization approaches struggle to handle the high-dimensional decision space and dynamic nature of these networks. This paper proposes a novel hierarchical deep reinforcement learning (HDRL) framework to address these challenges and enable intelligent spectrum sharing. The proposed framework leverages the inherent hierarchy of the network, with separate policies for each tier, to learn and optimize spectrum allocation decisions at different timescales and levels of abstraction. By decomposing the complex spectrum sharing problem into manageable sub-tasks and allowing for efficient coordination among the tiers, the HDRL approach offers a scalable and adaptive solution for spectrum management in future TN-NTNs. Simulation results demonstrate the superior performance of the proposed framework compared to traditional approaches, highlighting its potential to enhance spectral efficiency and network capacity in dynamic, multi-tier environments.
\end{abstract}

\section{Introduction}

The exponential growth of wireless data traffic and the emergence of new use cases have driven the need for integrating non-terrestrial networks (NTNs) with terrestrial networks (TNs) to meet the demanding requirements of future communications. NTNs, such as satellites, high-altitude platforms (HAPs), and unmanned aerial vehicles (UAVs), offer unique advantages in terms of coverage extension, service continuity, and rapid deployment. However, the increasing demand for wireless services has led to a scarcity of available spectrum resources. Spectrum sharing between NTNs and TNs has emerged as a promising solution to address this issue, enabling the efficient utilization of limited spectrum while accommodating the diverse requirements of future wireless networks.

The Third Generation Partnership Project (3GPP) has been actively working on incorporating NTN components into the 5G New Radio (NR) architecture, with ongoing studies and specifications in Release 17 and beyond~\cite{10417096,9221119}. The integration of NTNs with TNs introduces significant challenges in spectrum sharing and interference management due to the multi-tier, heterogeneous nature of the resulting network. The coexistence of diverse network elements with different characteristics, such as altitude, mobility, and transmission power, leads to a complex and dynamic interference landscape. Moreover, the high-dimensional decision space arising from the joint allocation of resources across multiple tiers renders conventional optimization approaches impractical for real-time spectrum management~\cite{electronics9091416}.

Recent work has explored various aspects of spectrum sharing in integrated TN-NTNs, including cognitive radio techniques~\cite{9686170}, dynamic spectrum access~\cite{10195296}, and interference mitigation schemes. However, these approaches often rely on simplifying assumptions and struggle to adapt to rapidly changing network conditions. Machine learning, particularly deep reinforcement learning (DRL), has emerged as a promising tool for complex resource management problems in wireless networks.

Several studies have applied DRL to spectrum sharing and resource allocation in NTNs, demonstrating its effectiveness in optimizing spectrum access and power control in cognitive satellite-terrestrial networks~\cite{10024896}, coordinating multiple HAPs for enhanced coverage~\cite{s22041630}, and managing interference between UAVs and ground users~\cite{10016705}. However, most existing DRL-based solutions focus on a single network tier or limited resource types, failing to capture the full complexity of the integrated TN-NTN environment. Cao et al.~\cite{10417096} proposed a multi-tier DRL approach for spectrum sharing in NTN networks, using separate policies for different network tiers. While this improves upon single-agent DRL by considering each tier's unique characteristics, it lacks an explicit hierarchical structure and the ability to handle multiple agents within each tier, which is crucial for efficient coordination and scalability in large-scale networks.

To address these limitations, we propose a novel hierarchical DRL (HDRL) framework for intelligent spectrum sharing in integrated TN-NTNs. Our approach mirrors the network's hierarchy to decompose the complex spectrum sharing problem into manageable sub-tasks, with meta-controllers guiding the learning of sub-controllers at each tier. By incorporating multiple policies within each tier and enabling coordination across tiers, the proposed framework offers a scalable and adaptive solution for dynamic spectrum management in integrated TN-NTNs.


\section{Spectrum Sharing in Integrated TN-NTNs}

In this section, we provide a brief overview of spectrum sharing challenges and opportunities in integrated TN-NTNs, highlighting the need for intelligent and adaptive spectrum management solutions.

\subsection{Existing Strategies and Advances}

Spectrum sharing in integrated TN-NTNs aims to efficiently utilize limited spectrum resources by allowing both TN and NTN components to access the same spectrum, either simultaneously or opportunistically. Traditional spectrum sharing techniques can be broadly classified into static and dynamic approaches. Static spectrum sharing involves pre-allocating fixed portions of the spectrum to different networks or users, often based on geographical separation or orthogonal frequency bands. In legacy systems, for instance, geostationary orbit (GEO) satellites and TNs operated in separate frequency bands. However, as spectrum demands grow, such fixed allocation becomes inefficient and leaves spectrum resources largely underutilized~\cite{10417096}.

Dynamic spectrum access (DSA) offers greater flexibility and adaptability. A common approach is opportunistic spectrum access, where secondary users (SUs) access licensed spectrum bands when not in use by primary users (PUs). SUs perform spectrum sensing to detect the presence or absence of PUs and adjust their transmission parameters accordingly. Techniques such as energy detection, matched filtering, and cyclostationary feature detection are commonly used in cognitive radio (CR) for this purpose. Another approach enables co-channel or adjacent-channel coexistence between networks, provided interference levels are constrained through techniques like power control, beamforming, and interference cancellation~\cite{10769081}. Recent advances have focused on making DSA more efficient, scalable, and adaptive to dynamic network conditions. For example, database-driven spectrum sharing uses a central database to store information about spectrum availability and usage rights, enabling more efficient coordination between users and networks~\cite{s23010342}. Advancements in CR systems have also contributed to the development of more intelligent DSA techniques; cooperative spectrum sensing allows SUs to collaborate and share sensing results to improve detection performance. Similarly, spectrum mobility techniques enable SUs to dynamically switch between different frequency bands or networks based on changing channel conditions or traffic demands.

\subsection{AI-Driven Spectrum Sharing}

While traditional DSA techniques improve upon static allocation, they often struggle with the complexity and dynamism of integrated TN-NTNs. Conventional optimization and game theory-based approaches typically require complete system information and rely on simplifying assumptions, making them impractical in real-world scenarios with limited information and heterogeneous network components. The emergence of AI-based solutions has opened new avenues for developing more intelligent spectrum sharing solutions. AI-driven techniques excel at learning from experience, adapting to changing network conditions, and making decisions in complex environments with incomplete information, thereby achieving efficient spectrum utilization and improved network performance.

Several AI techniques have been applied to spectrum sharing, including deep learning (DL), DRL, and federated learning (FL). DL models, such as convolutional neural networks and recurrent neural networks, are employed for tasks like spectrum sensing, channel prediction, and interference classification~\cite{10397567}. DRL algorithms enable autonomous agents to learn optimal spectrum access policies through environmental interaction and reward-based learning~\cite{10024896}. For instance, a multi-agent DRL-based scheme for spectrum sharing is proposed in~\cite{10817109}. The authors in~\cite{10791451} introduced a hierarchical spectrum sharing framework for cognitive satellite-terrestrial networks, enhancing spectrum efficiency through dynamic resource allocation. Furthermore, FL addresses privacy concerns and enables efficient, collaborative learning by allowing multiple devices to train a shared ML model without sharing their raw data, making it particularly valuable in distributed spectrum sharing scenarios.

\subsection{Challenges and Limitations}

Despite advances in spectrum sharing strategies, several limitations continue to hinder their full potential in integrated TN-NTNs. Traditional methods often rely on simplified assumptions about network characteristics and user behavior, which may not hold in real-world scenarios with heterogeneous network components and dynamic operating environments. Many recent methods assume perfect or near-perfect spectrum sensing capabilities, which is challenging to achieve in practice, especially in satellite communication where weak received signal strength and atmospheric effects significantly affect sensing accuracy. Additionally, centralized approaches may face scalability issues, particularly in large, geographically dispersed networks with numerous TN and NTN components.

AI-driven approaches also face challenges related to training data requirements, computational constraints, and explainability. The complexity of DL models and DRL agents can limit their deployment on resource-constrained devices, such as UAVs and satellites, which typically have limited processing capabilities and energy budgets. Furthermore, ensuring the explainability and interpretability of AI-based spectrum sharing decisions is crucial for gaining trust and facilitating regulatory approval. These limitations underscore the need for more intelligent, robust, and adaptable spectrum sharing frameworks for integrated TN-NTNs that can handle the challenges of complex, dynamic environments while considering practical constraints and addressing explainability concerns.

\section{Fundamentals of Hierarchical Deep Reinforcement Learning}

Before detailing the proposed framework for intelligent spectrum sharing in integrated TN-NTNs, we provide a general overview of DRL and its hierarchical extension, which forms the basis of our approach.

\subsection{Deep Reinforcement Learning}

DRL combines the function approximation capabilities of deep neural networks (DNNs) with the learning and decision-making framework of RL. This allows agents to learn complex policies directly from high-dimensional sensory inputs, such as images, sensor readings, or feedback signals, eliminating the need for manual feature engineering, though techniques like normalization and preprocessing remain beneficial.

\subsubsection{Markov Decision Process}

The mathematical framework underlying most RL algorithms is the Markov decision process (MDP). An MDP can be represented as a tuple $\langle S, A, P, R, \gamma \rangle$, where $S$ is the set of possible states, $A$ is the set of actions, $P$ is the state transition probability function, $R$ is the reward function, and $\gamma$ is the discount factor. $P(s'|s, a)$ defines the probability of transitioning to state $s'$ from state $s$ when action $a$ is taken. The reward function $R(s, a)$ specifies the immediate reward for taking action $a$ in state $s$ and plays a crucial role in shaping the agent's policy. The discount factor $\gamma \in [0, 1]$ determines the trade-off between immediate and future rewards. The goal is to learn a policy $\pi(a|s)$ that maximizes the expected cumulative discounted reward, known as the return~\cite{10016705}.

\begin{figure*}[t]
    \centering
    \includegraphics[width=0.8\linewidth]{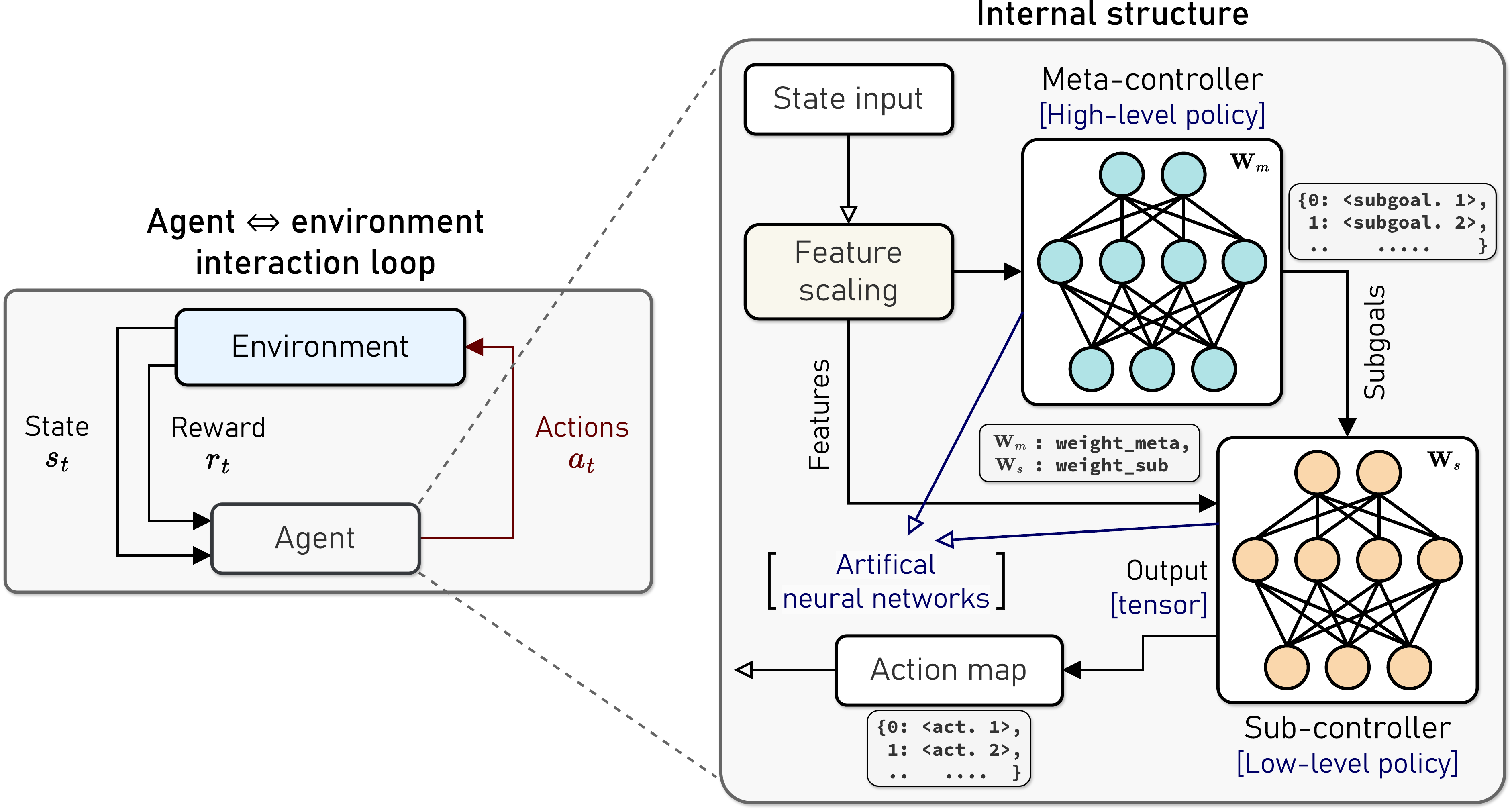}
    \caption{Illustration of the hierarchical policy structure and agent-environment interaction loop.}
    \label{fig:hrl_structure}
\end{figure*}

\subsubsection{DNNs as Function Approximators}

In many practical RL problems, the state and action spaces can be extremely large or continuous, making it infeasible to store and update Q-values~\footnote{Q-values represent the expected cumulative reward of taking an action $a$ in state $s$ and following a specific policy thereafter.} or policy probabilities for every state-action pair in a Q-table. DNNs, a class of artificial neural networks (ANNs) with multiple layers, serve as function approximators in DRL. According to the universal approximation theorem, DNNs can approximate any continuous function given sufficient capacity, allowing agents to learn complex mappings from states to actions or Q-values. The network takes the state as input and outputs either Q-values (in Q-learning) or the probability distribution over actions (in policy gradient methods). By training the DNN using appropriate algorithms and loss functions, the agent learns an effective policy to maximize its cumulative reward and optimize system control~\cite{10024896}.

\subsection{Hierarchical Deep Reinforcement Learning}

HDRL extends DRL by introducing hierarchies to the decision-making process, addressing challenges posed by complex real-world problems like spectrum sharing in integrated TN-NTNs. These problems often involve large action spaces that make learning slow and challenging. Instead of learning a single policy to map states to actions, HDRL agents learn multiple policies at different abstraction levels, decomposing the overall task into a hierarchy of sub-tasks. Each level of the hierarchy manages a different sub-task contributing to the core goal while maintaining scalability~\cite{3453160}. By decomposing complex problems into smaller, more manageable sub-problems, HDRL enables efficient learning and coordination among a large number of agents, making it inherently scalable to large-scale networks. The key principles of HDRL are as follows.

\begin{itemize}
    \item \textit{Hierarchical Policy:} High-level policies define abstract goals or sub-tasks, while low-level policies determine the specific actions to achieve them. This allows the agent to learn complex strategies by focusing on different levels of abstraction.

    \item \textit{Temporal Abstraction:} Control decisions occur at different timescales; high-level policies operate slower, setting long-term goals, while low-level policies act faster, refining actions based on immediate observations and higher-level directives. In context of an integrated TN-NTN scenario, a higher policy governed by a low Earth orbit (LEO) satellite might allocate spectrum resources on a larger, resource-constrained timescale, while lower policies for terrestrial nodes manage resource allocation within each region in real-time based on channel feedback and network load. This allows agents to plan and execute actions over varying time horizons, leading to more efficient learning and decision-making. Essentially, an agent can learn efficient strategies by avoiding constant decision-making for every aspect of the task.

    \item \textit{Sample Efficiency:} HDRL improves sample efficiency through its hierarchical structure and agents can learn from fewer environmental interactions. This is particularly valuable in scenarios where data collection is expensive or time-consuming.
\end{itemize}

Fig.~\ref{fig:hrl_structure} illustrates a simple hierarchical policy structure of an agent interacting with an environment. Although only two levels of hierarchy are shown, HDRL can be easily extended to multiple levels, each with its own policy and corresponding DNN. The agent interacts with the environment in a typical RL loop: it observes the state, selects an action based on the current policy, receives a reward, and updates the policy parameters. At the top level, a meta-controller outputs subgoals to the sub-controller, which in turn outputs actions to the action mapper. This hierarchical structure allows the agent to make decisions at different levels of abstraction and timescales, leading to more efficient and adaptive learning. Hierarchical control is particularly beneficial for integrated TN-NTNs, where network tiers with limited computational capabilities can use higher-level directives to simplify their decision-making processes~\cite{10417096}. HDRL agents can adapt to both long-term global changes and short-term local fluctuations, crucial for adaptive spectrum sharing in dynamic integrated TN-NTN environments.

\section{HDRL-Based Intelligent Spectrum Sharing}

In this section, we present our proposed HDRL-based framework for intelligent spectrum sharing in integrated TN-NTNs.

\subsection{System Model}

\begin{figure*}[t]
    \centering
    \includegraphics[width=0.98\linewidth]{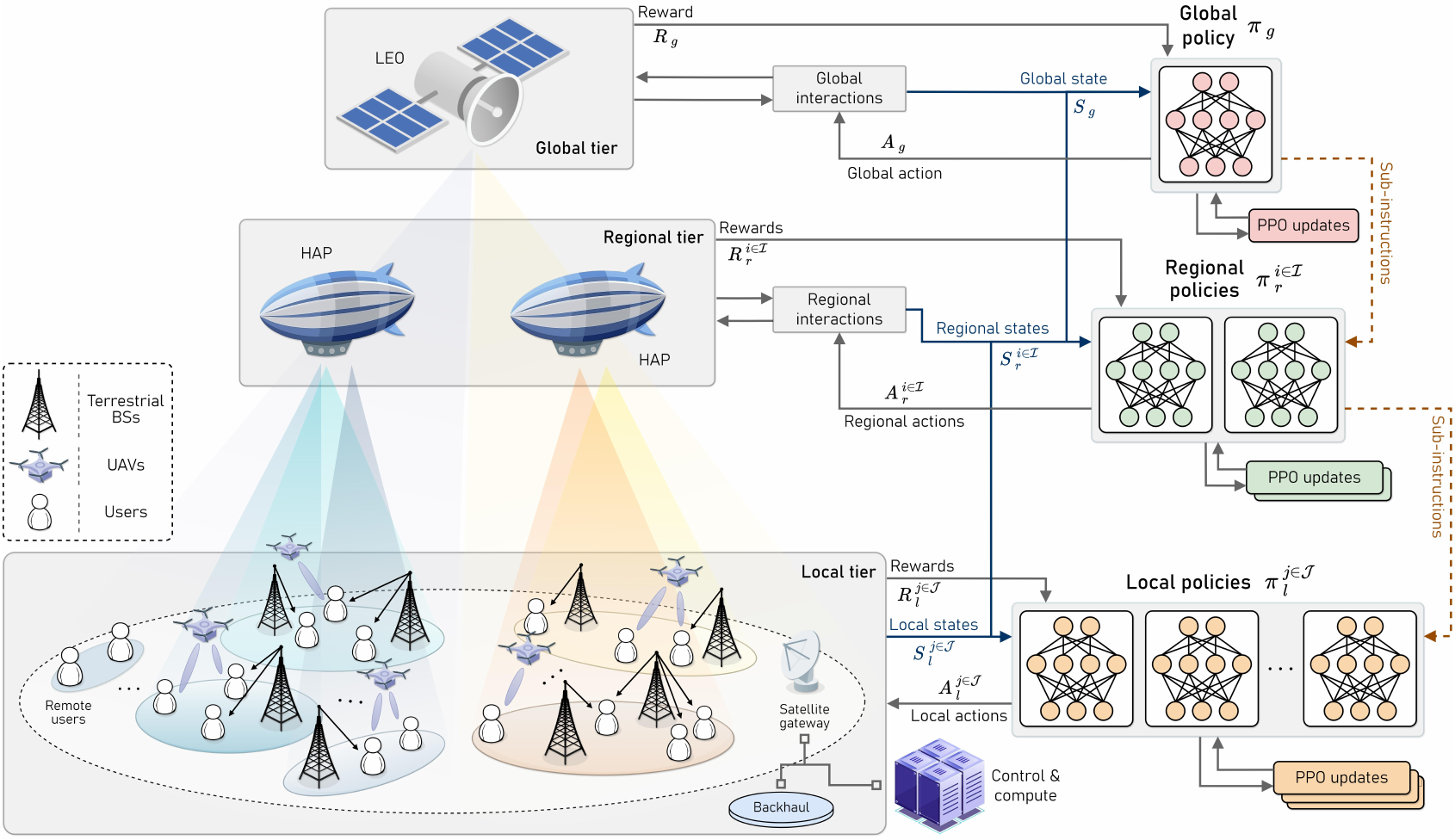}
    \caption{System model and HDRL framework for spectrum sharing in integrated TN-NTNs.}
    \label{fig:framework}
\end{figure*}

Fig.~\ref{fig:framework} illustrates our system comprising a LEO satellite, HAPs, UAVs, terrestrial base stations (TBSs), and users. The LEO satellite serves designated geographical areas using fixed multi-beam technology. A HAP acts as a regional hub in each beam cell, relaying data and control signals between the satellite and lower tiers. Multiple TBSs and UAVs operate within each HAP's coverage area. TBSs provide fixed high-capacity connectivity while UAVs function as aerial BSs offering flexible on-demand coverage. This setup addresses coverage gaps, enhances network capacity, and accommodates temporary hotspots or high-traffic events. Users dynamically associate with either a TBS or UAV based on signal strength, network load, and quality of service (QoS) requirements. A satellite gateway facilitates communication between the terrestrial network and the LEO satellite. A control and compute unit manages network operations as a central coordinator.

The LEO satellite allocates portions of its available spectrum to each beam cell, which the HAPs further divide and jointly access with UAVs and TBSs. To enhance spectral efficiency and user capacity, both UAVs and TBSs employ downlink non-orthogonal multiple access (NOMA). NOMA allows multiple users simultaneous access to the same frequency resource by exploiting power-domain multiplexing and successive interference cancellation (SIC) at the receiver~\cite{9520317}. To improve coverage, capacity, and cell-edge performance, UAVs and TBSs can utilize non-coherent joint transmission coordinated multi-point (JT-CoMP) in combination with NOMA (CoMP-NOMA), enabling multiple transmission points to cooperatively serve users.

This spectrum sharing scenario presents several key challenges: (1) maximizing overall network throughput through efficient inter-tier spectrum allocation; (2) managing inter-tier and intra-tier interference among nodes; and (3) adapting to the dynamic network environment including user mobility, varying channel conditions, and time-varying traffic demands. We propose an HDRL framework that enables intelligent and adaptive spectrum sharing across TN and NTN to address these challenges.

\subsection{Proposed Framework}

The proposed framework mirrors the network's hierarchical structure with three levels: global, regional, and local, as shown in Figure~\ref{fig:framework}. Each level corresponds to a specific network tier and employs a DNN as a policy to learn and optimize spectrum sharing via proximal policy optimization (PPO). Although we opt for PPO due to its stability and efficiency, the framework is adaptable to other RL algorithms such as soft actor-critic (SAC) or twin delayed deep deterministic policy gradient (TD3) as well.

Unlike single-agent PPO, which optimizes a single policy, or multi-agent PPO (MAPPO), where each agent optimizes its own policy independently, our framework utilizes a hierarchical approach. Higher-level policies provide context or subgoals for lower-level policies, constraining their action spaces. This enables more efficient learning and decision-making in the complex spectrum sharing environment. The specific roles of each tier are as follows.

\begin{itemize}
    \item \textit{Global Tier:} The global policy $\pi_g$ oversees the overall spectrum allocation for the entire network. It determines the optimal distribution of total spectrum chunks across beam cells based on the global network state $S_g$, which includes aggregated information about spectrum demand, user distribution, and channel conditions. This high-level allocation constrains the regional tier policies, ensuring fair and efficient spectrum distribution across different geographical areas. The reward function for the global tier, denoted as $R_g$, maximizes overall spectral efficiency while considering fairness among beams and adherence to QoS requirements. It is defined as
    \begin{equation}
        R_g = w_{1}^{(g)} E_g + w_{2}^{(g)} F_g - w_{3}^{(g)} V_g,
    \end{equation}
    where $E_g$ is the normalized spectral efficiency of the satellite, $F_g$ represents Jain's fairness index for spectrum allocation, and $V_g$ denotes QoS violations across all beams. The weights $w_{1}^{(g)}$, $w_{2}^{(g)}$, and $w_{3}^{(g)}$ balance the trade-offs between spectral efficiency, fairness, and QoS.

    \item \textit{Regional Tier:} Each HAP employs a regional policy $\pi_r^i$ to manage spectrum resources within its designated coverage area. Given the spectrum allocation from the global tier, the regional policy divides and assigns spectrum sub-bands to UAVs and TBSs under its control. This assignment relies on the regional network state $S_r^i$, which incorporates information about user distribution, traffic demands, and channel conditions within the HAP's coverage region. The regional policy adapts its sub-band assignments to optimize regional performance while adhering to the global tier's allocation. The reward function for the regional tier, $R_r^i$, is expressed as
    \begin{equation}
        R_r^i = w_{1}^{(r)} E_r^i + w_{2}^{(r)} F_r^i - w_{3}^{(r)} V_r^i
    \end{equation}
    where each term mirrors the global reward function but applies specifically to region $i$. The weights in this tier are chosen to align regional objectives with the global objective.

    \item \textit{Local Tier:} UAVs and TBSs utilize local policies to control real-time spectrum access and power allocation for their associated users. For each node $j$, the local policy $\pi_l^j$ makes fine-grained decisions on specific channels and power levels for each connected user, based on the local network state $S_l^j$. This state includes information about individual user requirements, channel gains, and interference levels. The local policies operate within the spectrum sub-bands assigned by the regional tier, ensuring compliance with higher-level decisions while adapting to local network dynamics. A similar reward function $R_l^j$ applies to each local policy, with an additional penalty term for UAVs' mobility to prevent them from leaving their designated coverage area.
\end{itemize}

It should be noted that the specific weight values are obtained through hyperparameter tuning via grid search, a common practice in training DRL agents to optimize overall performance. This hierarchical structure enables decision-making at different timescales and abstraction levels, aligning with each tier's unique operational characteristics and requirements.


\subsubsection{Learning Process}

The learning process in the proposed HDRL framework involves interactions between policies at different tiers and the environment. Initially, each policy is randomly initialized. The process begins with the local tier sensing their environment and constructing local state representations. These local states are then aggregated at the regional tier to form regional state representations, which incorporate information from the lower tier and the global spectrum allocation. At the highest level, i.e., global tier, aggregated information from all HAPs is received forming the global state.

Based on these state representations, each tier executes actions according to its current policy, with higher-tier actions serving as subgoals or constraints for lower-tier policies. Each policy is updated based on the received reward and the observed state transitions using the PPO algorithm. This iterative process of interaction, reward collection, and policy update continues until the learning converges to a near-optimal solution. While neural networks inherently possess some ability to adapt to changing environments, significant changes in the network structure or operating conditions may require retraining the policies. Techniques like transfer learning and meta-learning can expedite this process by utilizing previously learned knowledge. The exact frequency of retraining, however, depends heavily on a practical deployment scenario and evaluation of trade-off between computational cost and adaptability.

\subsubsection{Complexity Analysis}

The computational complexity of the proposed HDRL framework is influenced by the decision space dimension and learning complexity at each tier. The decision space dimension $|\mathcal{D}|$ represents the number of possible action combinations a policy can choose from. At the global tier, the satellite allocates spectrum chunks to $B$ beam cells from a pool of $F$ available frequency bands. When beam cells can receive multiple frequency bands, the corresponding space is $|\mathcal{D}_g| = \binom{F+B-1}{B}$. For cases where each beam cell receives only one frequency band and frequency bands can be allocated to multiple beam cells, the decision space dimension becomes $|\mathcal{D}_g| = \sum_{k=1}^{\min(F,B)} S(B,k)\binom{F}{k}$, where $S(B,k)$ represents the Stirling numbers of the second kind. At the regional tier, each HAP divides its allocated spectrum into $S$ sub-bands for $N_u$ UAVs and $N_t$ TBSs. Assuming each node can be allocated multiple sub-bands, the decision space dimension for each HAP is $|\mathcal{D}_r| = S^{(N_u + N_t)}$. At the local tier, each UAV/TBS controls spectrum access and power allocation for $M$ associated users by selecting from $C$ channels and $P$ power levels. The decision space for each local policy is $|\mathcal{D}_l| = (C \times P)^M$. The overall decision space complexity can be approximated as the product of the decision space dimensions across all tiers: $|\mathcal{D}| = |\mathcal{D}_g| \times |\mathcal{D}_r| \times |\mathcal{D}_l|$. The hierarchical structure significantly reduces this complexity compared to a flat, non-hierarchical approach, as higher-tier actions constrain the decision space of lower tiers~\cite{10417096}.

In terms of training complexity, each epoch involves forward and backward passes through the neural networks at each tier alongside policy updates. For a network with $L$ layers and $H$ hidden units per layer, each forward/backward pass has a complexity of $O(LH^2)$. The total training complexity per epoch for our hierarchical framework, considering $R$ regions per HAP and $N$ number of TBSs and UAVs, is approximately $O(B_g LH_g^2 + |\mathcal{D}_g| + R(B_r LH_r^2 + |\mathcal{D}_r|) + N(B_l LH_l^2 + |\mathcal{D}_l|))$, where $B_g$, $B_r$, and $B_l$ are the batch sizes and $H_g$, $H_r$, and $H_l$ are the hidden units of the global, regional, and local policies respectively. During inference, the complexity is dominated by forward passes, resulting in a total execution time complexity of $O(H_g^2 + |\mathcal{D}_g| + R(H_r^2 + |\mathcal{D}_r|) + N(H_l^2 + |\mathcal{D}_l|))$.

\section{Performance Evaluation}

\begin{figure}[!t]
    \centering
    \includegraphics[width=0.985\linewidth]{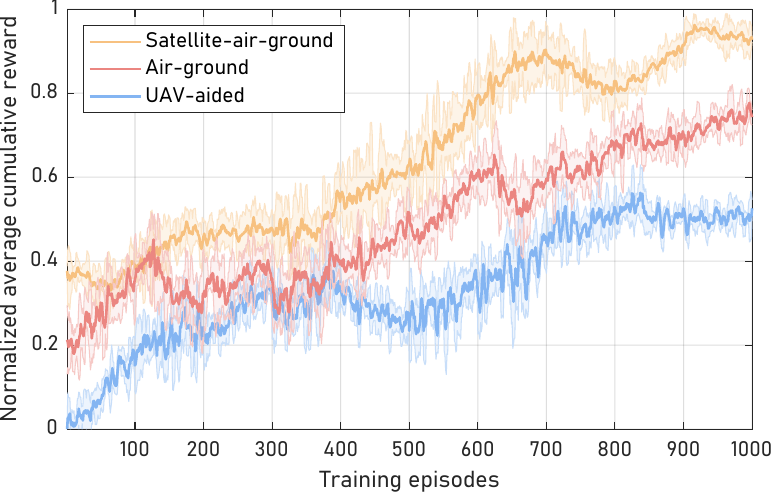}
    \caption{Normalized average cumulative reward of the proposed HDRL framework for different network hierarchies.}
    \label{fig:convergence}
\end{figure}

\begin{figure}[!t]
    \centering
    \includegraphics[width=0.985\linewidth]{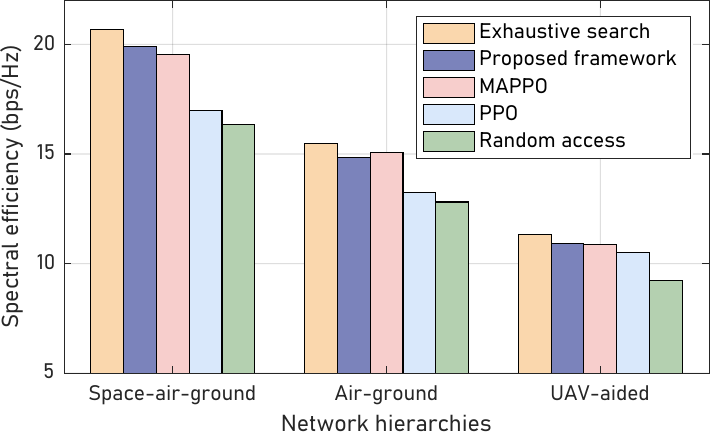}
    \caption{Spectral efficiency achieved by different algorithms for different network hierarchies.}
    \label{fig:spectral}
\end{figure}

\begin{figure}[!t]
    \centering
    \includegraphics[width=0.985\linewidth]{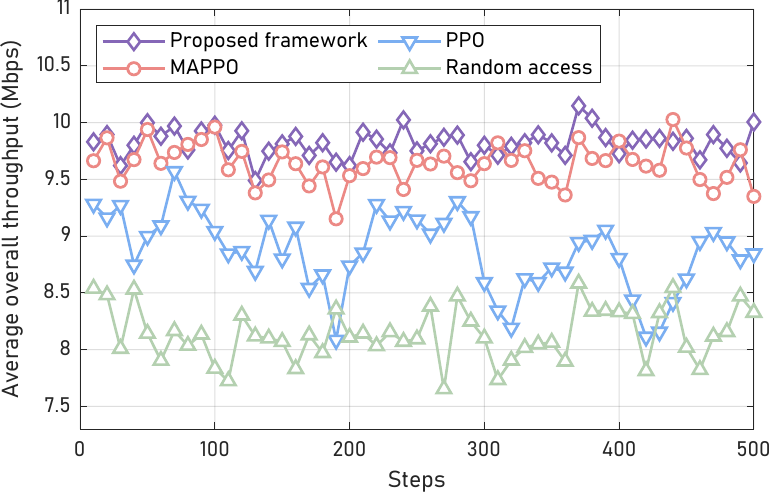}
    \caption{Average network throughput achieved by different algorithms.}
    \label{fig:throughput}
\end{figure}

We now evaluate the performance of the proposed HDRL-based spectrum sharing framework for integrated TN-NTNs through extensive simulations. The proposed approach is compared with several baseline methods, including exhaustive search, random access, and single- and multi-agent RL algorithms.

\subsection{Simulation Setup}

All experiments were conducted on a system equipped with an Intel Xeon Platinum 8259CL CPU, 32GB of RAM, and an NVIDIA Tesla T4 GPU with 16GB VRAM. We simulate a system similar to that shown in Fig.~\ref{fig:framework}. Specifically, the LEO satellite operates at an altitude of $550$ km with two fixed beams, using Ka-band at $28$ GHz with a total bandwidth of $200$ MHz. Within each beam, a HAP is deployed at an altitude of $20$ km each cover two non-overlapping regions. These regions accommodate two TBSs and one UAV that jointly serve terrestrial users. TBSs have high-power transmitters, while UAVs operate at an altitude of $100$ m with lower transmission power, offering flexible deployment to enhance coverage and capacity.

The number of users per HAP region varies from $10$ to $30$, following a uniform distribution. Users dynamically associate with either TBSs or UAVs based on signal strength, network load, and QoS requirements. The LEO satellite dynamically allocates portions of its $200$ MHz bandwidth to HAPs based on demand and network conditions. HAPs further divide their allocated spectrum into $10$ sub-bands for TBSs and UAVs, enabling efficient resource utilization across the network tiers.

To capture the diverse propagation characteristics, we employ a shadowed Rician fading model for satellite links and Rayleigh fading for terrestrial links. The channel models incorporate path loss, atmospheric effects, and small-scale fading appropriate for each link type. The simulation time is divided into discrete slots. The satellite makes decisions every $50$ slots, HAPs every $10$ slots, and TBSs/UAVs every slot, reflecting the different computational capabilities and control cycles of network entities. The SINR for each user is calculated considering intra-tier and inter-tier interference, as well as noise. Throughput is derived using the Shannon capacity formula: $C = B \log_2(1 + \text{SINR})$, where $B$ is the bandwidth and SINR is the signal-to-interference-plus-noise ratio.

\begin{table}[t!]
\caption{Comparison of execution time for different algorithms.}
\label{tab:execution_time}
\centering
\resizebox{0.985\columnwidth}{!}{
\begin{tabular}{l c c c}
    \toprule
    \textbf{Algorithm} & \multicolumn{3}{c}{\textbf{Execution time (ms)}} \\
    \cmidrule(lr){2-4}
    & Space-air-ground & Air-ground & UAV-aided \\
    \midrule
    Exhaustive search & 420.3 & 371.7 & 320.8 \\
    Random access & 0.37 & 0.36 & 0.36 \\
    PPO & 1.40 & 1.25 & 1.10 \\
    MAPPO & 31.1 & 28.4 & 26.3 \\
    \textbf{Proposed framework} & 7.61 & 6.89 & 6.60 \\
    \bottomrule
\end{tabular}
}
\end{table}

\subsection{Results and Discussion}

We compare the performance of our proposed HDRL framework against several baseline approaches:

\begin{itemize}
    \item \textit{Exhaustive Search}: This method explores all possible spectrum allocation combinations to find the optimal solution. It guarantees the best performance but is computationally intensive and infeasible for large-scale deployments.
    \item \textit{Random Access}: This approach randomly allocates spectrum resources, serving as a lower bound for performance.
    \item \textit{PPO}: A single-agent RL algorithm optimizing a single policy for the entire network.
    \item \textit{MAPPO}: A multi-agent version of PPO where each network node optimizes its own policy independently.
\end{itemize}

Unless stated otherwise, we consider space-air-ground network as the default scenario. The proposed framework is evaluated in terms of learning and convergence, spectral efficiency, average network throughput, and execution time. For training the PPO agents, we used the following hyperparameters: learning rate of 0.0005, train batch size of 2000, stochastic gradient updates, discount factor $\gamma$ of 0.99, clip parameter of 0.2, entropy coefficient of 0.01, and a value function loss coefficient of 1.0. Convergence was determined when the improvement in normalized average reward was less than 0.15 for 10 consecutive trraining episodes.

\subsubsection{Learning and Convergence}

Fig.~\ref{fig:convergence} illustrates the learning and convergence behavior of our framework across satellite-air-ground, air-ground, and UAV-aided network hierarchies. The average cumulative reward, normalized for fair comparison, is plotted against the number of training episodes. For each hierarchy, the rewards steadily increase over time, converging to a stable value around $700$ to $800$ episodes. However, the rewards exhibit fluctuations around the converged value due to the dynamic nature of the channels and evolving network conditions. This result demonstrates the ability of HDRL to learn effective spectrum sharing policies in various network configurations while adapting to the specific challenges and opportunities presented by each hierarchy.

\subsubsection{Spectral Efficiency}

Fig.~\ref{fig:spectral} compares the spectral efficiency achieved by different spectrum sharing algorithms across the three network hierarchies. Our hierarchical framework achieves near-optimal performance, as compared to exhaustive search, demonstrating its effectiveness in optimizing spectrum utilization. MAPPO also achieves similar performance and even outperforms our framework in air-ground and UAV-aided networks, which can be attributed to simpler network structures and fewer nodes. PPO struggles due to the large action space and lack of coordination among network nodes, resulting in suboptimal performance. The random access scheme yields the lowest spectral efficiency since it allocates resources without considering network conditions or user requirements. These results highlight the advantages of our hierarchical approach in efficiently managing spectrum resources while maintaining scalability and adaptability.

\subsubsection{Average Network Throughput}

Fig.~\ref{fig:throughput} presents a comparison of the average network throughput achieved by various algorithms over the course of an episode, with measurements taken every $10$ steps. Our hierarchical framework consistently achieves better performance than other algorithms. MAPPO performs similarly to our framework but exhibits higher variance, likely due to autonomous agents making independent decisions without explicit coordination despite being in a collaborative setting. PPO yields the lowest throughput among the learning-based methods, as it struggles to handle the exploded action space and lacks coordination among network nodes. These findings highlight the importance of leveraging the network's hierarchical structure and enabling efficient inter-tier collaboration for maximizing system performance.

\subsubsection{Execution Time}

Table~\ref{tab:execution_time} compares the execution time of different spectrum sharing algorithms across all network hierarchies. The execution time is calculated by averaging the time taken by each control node to make a decision across all steps in an episode, providing a fair comparison considering the varying control cycles of the global, regional, and local tiers in our hierarchical framework. The random access scheme exhibits the lowest execution time since it does not involve any learning or optimization processes. The exhaustive search method requires the most time due to its loop-based exploration of all possible action combinations. Among the learning-based approaches, PPO achieves a low execution time, despite its suboptimal performance, as it optimizes a single policy for the entire network. Our hierarchical framework outperforms the MAPPO algorithm in terms of execution time, benefiting from the fact that not all policies are executed at each step, unlike the independent agents in the multi-agent setting. It can also be observed that the execution time decreases as the complexity of the network hierarchy decreases (from space-air-ground to UAV-aided), as fewer tiers and agents are involved in the decision-making process, leading to a reduction in the computational burden.

\section{Conclusion and Future Directions}

This article proposed an HDRL-based framework for intelligent spectrum sharing in integrated TN-NTNs. The framework leverages the network's inherent hierarchy, with separate policies for each tier, to learn and optimize spectrum allocation decisions at different timescales and levels of abstraction. By decomposing the complex spectrum sharing problem into manageable sub-tasks and enabling efficient inter-tier coordination, the HDRL approach offers a scalable and adaptive solution for spectrum management in future TN-NTNs. Simulation results demonstrate the performance gains of our framework compared to other approaches, such as exhaustive search, random access, and other DRL frameworks. HDRL achieves near-optimal spectral efficiency and network throughput while maintaining low execution times, making it suitable for real-time applications in future 6G networks.



\vskip -2\baselineskip plus -1fil
\begin{IEEEbiographynophoto}{Muhammad Umer} (\href{mailto:mumer.bee20seecs@seecs.edu.pk}{mumer.bee20seecs@seecs.edu.pk}) received the B.E. degree in electrical engineering from National University of Sciences and Technology (NUST), Pakistan. His current research interests include coordinated multi-point (CoMP) transmission, non-orthogonal multiple access (NOMA), reconfigurable intelligent surface (RIS), and deep reinforcement learning (DRL).
\end{IEEEbiographynophoto}

\vskip -2\baselineskip plus -1fil
\begin{IEEEbiographynophoto}{Muhammad Ahmed Mohsin} (\href{mailto:mmohsin.bee20seecs@seecs.edu.pk}{mmohsin.bee20seecs@seecs.edu.pk}) received the B.E. degree in electrical engineering from National University of Sciences and Technology (NUST), Pakistan. He is currently pursuing the Ph.D. degree in electrical engineering from Stanford University, USA. His primary focus of research lies in next-generation wireless communications, deep learning (DL), and reinforcement learning (RL).
\end{IEEEbiographynophoto}

\vskip -2\baselineskip plus -1fil
\begin{IEEEbiographynophoto}{Ali Arshad Nasir} (\href{mailto:anasir@kfupm.edu.sa}{anasir@kfupm.edu.sa}) received the Ph.D. degree in electrical engineering from Australian National University (ANU), Canberra, in 2013. He is currently an Associate Professor with the Department of Electrical Engineering, King Fahd University of Petroleum and Minerals (KFUPM), Dhahran, Saudi Arabia. His research interests are in the area of signal processing in wireless communication systems.
\end{IEEEbiographynophoto}

\vskip -2\baselineskip plus -1fil
\begin{IEEEbiographynophoto}{Hatem Abou-Zeid} (\href{mailto:hatem.abouzeid@ucalgary.ca}{hatem.abouzeid@ucalgary.ca}) received the Ph.D. degree in electrical and computer engineering from Queens University at Kingston. He is currently an Assistant Professor with the Schulich School of Engineering, University of Calgary. His research expertise is in communication networks, artificial intelligence (AI), and beyond 5G applications. Within these areas, he focuses on foundational models and trustworthy AI for 6G, wireless sensing, extended reality networking, and AI-powered brain-computer interfaces.
\end{IEEEbiographynophoto}

\vskip -2\baselineskip plus -1fil
\begin{IEEEbiographynophoto}{Syed Ali Hassan} (\href{mailto:ali.hassan@seecs.edu.pk}{ali.hassan@seecs.edu.pk}) received the Ph.D. degree in electrical engineering from Georgia Tech, Atlanta, in 2011. He is currently a Professor with the School of Electrical Engineering and Computer Science (SEECS), NUST, where he is also the Director of the Information Processing and Transmission Research Group, which focuses on various aspects of theoretical communications.
\end{IEEEbiographynophoto}

\end{document}